\documentstyle[12pt]{article}
\title{Population dynamics with a stable efficient equilibrium} 
\author{Jacek Mi\c{e}kisz and Tadeusz P\l atkowski
\\ Institute of Applied Mathematics \\
and Mechanics \\ Warsaw University  \\ ul. Banacha 2  \\ 02-097
Warsaw, Poland} 
\begin{document} 
\baselineskip=20pt
\maketitle 

\noindent {\bf Abstract}: 
We propose a game-theoretic dynamics of a population of replicating individuals. 
It consists of two parts: the standard replicator one and a migration between 
two different habitats. We consider symmetric two-player games 
with two evolutionarily stable strategies: the efficient one 
in which the population is in a state with a maximal payoff 
and the risk-dominant one where players are averse to risk. 
We show that for a large range of parameters of our dynamics,
even if the initial conditions in both habitats are in the basin 
of attraction of the risk-dominant equilibrium (with respect 
to the standard replication dynamics without migration), 
in the long run most individuals play the efficient strategy.
\vspace{3mm}

\vspace{3mm}

\noindent Corresponding author: Jacek Mi\c{e}kisz, e-mail: miekisz@mimuw.edu.pl, 
phone: 48-22-5544423, fax: 48-22-5544300.
\eject

\section{Introduction}

\newtheorem{theo}{Theorem}
\newtheorem{defi}{Definition}
\newtheorem{hypo}{Hypothesis}

The long-run behavior of interacting individuals can often be described within 
game-theoretic models. Players with different behaviors (strategies) receive payoffs 
interpreted as the number of their offspring. Maynard Smith and Price (1973) 
(see also Maynard Smith, 1982) introduced the fundamental notion of evolutionarily stable strategy. 
If everybody plays such a strategy, then a small number of mutants playing 
a different one is eliminated from the population. 
The dynamical interpretation of an evolutionarily stable strategy was later provided 
by several authors (Taylor and Jonker, 1978; Hofbauer {\em et al.}, 1979; Zeeman, 1981). 
They proposed a system of differential equations, the so-called replicator 
equations, which describe time-evolution of frequencies of strategies. 
It is known that any evolutionarily stable strategy is an asymptotically 
stable stationary point of such dynamics (Weibull, 1995; Hofbauer and Sigmund, 1998 and 2003).
 
We analyze symmetric two-player games with two evolutionarily stable pure strategies 
and a unique unstable mixed Nash equilibrium. The efficient strategy (called also payoff dominant), 
when played by the whole population, results in the highest possible payoff (fitness).
The risk-dominant one is played by individuals averse to risk. The strategy is risk-dominant 
if it has a higher expected payoff against a player playing both strategies 
with equal probabilities than the other one (the notion of the risk-dominance 
was introduced and thoroughly studied by Hars\'{a}nyi and Selten (1984)).

Typical example is that of a modified stag-hunt game, where two players choose simultaneously 
one of two actions: either to join the stag hunt (strategy $S$) or to go after a hare (strategy $H$).
The strategy $S$  gives the highest possible payoff provided  both players join the stag hunt 
and split the reward. However, when the other player is not loyal and switches to $H$, then
the one playing $S$ receives nothing. The strategy $H$ yields lower but less risky payoffs
with a higher payoff when the other player also chooses $H$.
Such a coordination game has two pure Nash equilibria: 
the efficient one, where all players choose $S$, and the one in which players 
are averse to risk, and play $H$. Both strategies are evolutionarily stable 
and therefore we are faced with the problem of equilibrium selection - which 
strategy will be played in the long run? 

In the replicator dynamics, both strategies are locally asymptotically stable, 
with the risk-dominant one having a bigger basin of attraction. Here we propose 
a dynamical model for which the efficient strategy has a much larger basin of attraction 
than the risk-dominant one. 

We consider a large population of identical individuals who at each time step 
can belong to one of two different nonoverlapping subpopulations or habitats 
which differ only by their replication rates. In both habitats they play 
the same two-player symmetric game. Our population dynamics consists of two parts: 
the standard replicator one and a migration between subpopulations. Individuals are allowed 
to change their habitats. They move to a habitat in which the average payoff 
of their strategy is higher; they do not change their strategies.

Migration can help the population to evolve towards an efficient equilibrium.
Below we briefly describe the possible mechanism responsible for it. 
If in a subpopulation, the fraction of individuals playing the efficient strategy $A$ 
is above its unique mixed Nash equilibrium fraction, then the expected payoff of $A$ 
is bigger than that of $B$ in this subpopulation, and therefore the subpopulation evolves 
to the efficient equilibrium by the replicator dynamics without any migration. 
Let us assume therefore that such fraction is below the Nash equilibrium in both subpopulations. 
Without loss of generality we assume that initial conditions are such that the fraction 
of individuals playing $A$ is bigger in the first subpopulation than in the second one. 
Hence the expected payoff of $A$ is bigger in the first subpopulation than in the second one, 
and the expected payoff of $B$ is bigger in the second subpopulation than in the first one. 
This implies that a fraction of $A$-players in the second population will switch to the first one 
and at the same time a fraction of $B$-players from the first population will switch to the second one - 
migration causes the increase of the fraction of individual of the first population playing $A$. 
However, any $B$-player will have more offspring than any $A$-player 
(we are below a mixed Nash equilibrium) and this has the opposite effect on relative number 
of $A$-players in the first population than the migration. The asymptotic composition 
of the whole population depends on the competition between these two processes.

In this note we derive sufficient conditions for migration and replication rates such that
the whole population will be in the long run in a state in which most individuals 
occupy only one habitat (the first one for the above described initial conditions) 
and play the efficient strategy.

In Section 2, we introduce the notation and present the class of games we consider.
In Section 3, we propose a discrete-time model, obtain its continuous counterpart and prove our results. 
Section 4 contains a short discussion. 
\eject

\section{Replicator dynamics}
We consider two-player symmetric games with two pure strategies and two symmetric 
Nash equilibria. A payoff matrix is given by
\vspace{3mm}

\hspace{23mm} A  \hspace{2mm} B   

\hspace{15mm} A \hspace{3mm} a  \hspace{3mm} b 

U = \hspace{6mm} 

\hspace{15mm} B \hspace{3mm} c  \hspace{3mm} d,

where the $ij$ entry, $i,j = A, B$, is the payoff of the first (row) player when
he plays the strategy $i$ and the second (column) player plays the strategy $j$; 
payoffs of the column player are given by the matrix transposed to $U$. 

An assignment of strategies to both players is a Nash equilibrium, if for each player, 
for a fixed strategy of his opponent, changing the current strategy cannot increase his payoff.

In order to use our payoff matrix in the dynamics of reproducing species, 
we assume that all payoffs are non-negative. We also assume that $a>d>c$, 
$d>b$, and $a+b<c+d$ which implies in particular that $(A,A)$ and $(B,B)$ 
are two strict Nash equilibria hence both $A$ and $B$ are evolutionarily stable. 
The last inequality means that an expected payoff of $B$ against a player playing both strategies 
with equal probabilities is higher than that of $A$. We address the problem of equilibrium selection 
between the payoff-dominant (efficient) equilibrium $(A,A)$ and the risk dominant $(B,B)$. 
Our assumptions imply also that a $B$-player receives a maximal payoff when he plays against 
another $B$-player.  

As an example, consider the following payoffs: $a=4, b=0, c=2, d=3.$
The strategy profile $(A,A)$ is more risky than $(B,B)$ since from the point 
of view of the row player, a deviation by the column player in $(A,A)$ 
results in a payoff loss of $4$ units versus a loss of $1$ in $(B,B)$.

In the replicator dynamics, individuals of a large population are matched many times to play 
the above described stage game. Let $x$ be the fraction of the population playing $A$. 
The expected payoff of $A$ is given by $f_{A}=ax+b(1-x)$ and that of $B$ by $f_{B}=cx+d(1-x)$. 
A mixed Nash equilibrium strategy is such $x^{*}$ that the above two expected values are equal, hence
$x^{*}= (d-b)/(d-b+a-c).$ In the standard replicator dynamics, the rate of change of $x$ 
is proportional to the expected payoff of $A$. The corresponding differential equation reads 
(Weibull, 1995; Hofbauer and Sigmund, 1998)

\begin{equation}
\frac{dx}{dt} = x(f_{A}-f_{B})=x(1-x)(x-x^{*}). 
\end{equation}

If the initial condition $x(0)<x^{*}$, then the population evolves in time to the state $x=0$,
i.e., in the long run almost all individuals will play $B$. If $x(0)> x^{*}$, then $x=1$ 
is an attracting point of the dynamics. If $B$ is risk-dominant, then $x^{*}>1/2$, 
hence $B$ has a bigger basin of attraction than $A$. 

Below we propose an evolutionary dynamics for which an efficient strategy 
has a much bigger basin of attraction than the risk-dominant one. 

\section{Mixed replicator dynamics with migration}

We consider a large population of identical individuals who at each time step 
can belong to one of two different subpopulations or habitats. In both subpopulations 
they play the same two-player symmetric game.
Our population dynamics consists of two parts: the standard replicator one 
which represents slow evolutionary changes and a migration between two habitats.
Let $\delta$ be a time step. We allow two subpopulations 
to replicate with different speeds. We assume that during time-step $\delta$, 
a fraction $\delta$ of the first subpopulation and a fraction $\kappa\delta$ 
of the second subpopulation plays the game and receives payoffs 
which are interpreted as the number of their offspring. Moreover, we allow 
a fraction of individuals to migrate to a habitat in which their strategies 
have a higher expected payoff. 

Let $r^i_s$ denote the number of individuals which use the strategy $s \in \{A, B\}$ 
in the subpopulation $i \in \{1, 2\}$. By $f^i_s$ we denote the expected payoff 
of the strategy $s$ in the subpopulation $i$:  

$$ f^1_A = ax + b(1-x), \quad   f^1_B = cx + d(1-x),$$
$$ f^2_A = ay + b(1-y), \quad   f^2_B = cy + d(1-y),$$

where
$$ x= {r^1_A\over r_1}, \quad y = {r^2_A\over r_2},  \quad r_1 = r^1_A + r^1_B,  
\quad   r_2 = r^2_A + r^2_B;$$
$x$ and $y$ denote fractions of $A$-players in the first and second population
respectively. We denote by $\alpha = {r_1\over r}$ the fraction of the whole population 
in the first subpopulation, where $r = r_1 + r_2$ is the total number of individuals.
 
The evolution of the number of individuals in each subpopulation is assumed to be a result 
of the replication and the migration flow. 
In our model, the direction and intensity of migration of individuals with a given strategy 
will be determined by the difference of the expected payoffs of that strategy in both 
habitats. Individuals will migrate to a habitat with a higher payoff. 
The evolution equations for the number of individuals playing 
strategy $s$, $s \in \{A, B\}$, in habitat $i$, $i \in \{1, 2\}$, have the following form: 

\begin{equation}
r^1_A(t+\delta) = R^1_A + \Phi_A,
\end{equation}
\begin{equation}
r^1_B(t+\delta) = R^1_B + \Phi_B, 
\end{equation}
\begin{equation}
r^2_A(t+\delta) = R^2_A - \Phi_A,
\end{equation}
\begin{equation}
r^2_B(t+\delta) = R^2_B - \Phi_B,
\end{equation}
where all functions on the right-hand sides are calculated at time $t$.

Functions $R^i_s$ describe an increase of the number of the individuals playing 
strategy $s$ in the subpopulation $i$ due to replication, cf. (Weibull, 1995): 

\begin{equation}
R^1_s= (1-\delta) r^1_s 
+ \delta f^1_s r^1_s, \ \ s \in \{A, B\},
\end{equation}
\begin{equation}
R^2_s= (1-\kappa \delta) r^2_s  
+ \kappa \delta f^2_s r^2_s, \ \ s \in \{A, B\},
\end{equation}

The rate of replication of the individuals playing strategy $s$ 
in the first subpopulation is given by $\delta f^1_s$ , 
and in the second subpopulation by $\kappa \delta f^2_s$. 
The parameter $\kappa$ measures the difference of reproduction speeds 
in both habitats.   

Functions $\Phi_s$, $s \in \{A, B\}$, are defined by 
\begin{equation}
\Phi_s = \delta \gamma (f^1_s - f^2_s)[r^2_s \Theta (f^1_s - f^2_s) + r^1_s \Theta (f^2_s-f^1_s)], 
\end{equation} 
where $\Theta$ is the Heaviside's function, 
\begin{equation}
\Theta(x) = \cases {1, \ \ x\ge 0; \cr 0, \ \ x<0.\cr}
\end{equation}

Functions $\Phi_s$ describe changes of the numbers of the individuals 
playing strategy $s$ in the relevant habitat due to migration. 
$\Phi_s$ will be referred to as the migration 
of individuals (who play strategy $s$) between the habitats. 

Thus, if for example $f^1_A > f^2_A$, then there is a migration of individuals with 
strategy $A$ from the second habitat to the first one: 
\begin{equation}
\Phi_A = \delta \gamma r^2_A (f^1_A - f^2_A),
\end{equation}
and since then necessarily $f^1_B < f^2_B$ [note that 
$f^1_A - f^2_A = (a-b)(x-y) \hspace{3mm} and  \hspace{3mm} f^1_B - f^2_B = (c-d)(x-y)$], 
there is a migration flow of individuals with strategy $B$ from the first habitat to the second one: 
\begin{equation}
\Phi_B = \delta \gamma r^1_B (t) (f^1_B - f^2_B). 
\end{equation}

In this case, the migration flow $\Phi_A$ describes the increase of the number of individuals 
which play strategy $A$ in the first subpopulation due to migration 
of the individuals playing strategy $A$ in the second subpopulation.
This increase is assumed to be proportional to the number of individuals playing strategy $A$ 
in the second subpopulation and the difference of payoffs of this strategy in both 
subpopulations. The constant of proportionality is $\delta$ times the rate of migration $\gamma$. 

The case $\gamma=0$ corresponds to two separate populations which do not communicate 
and evolve independently. Our model reduces then to the standard discrete-time replicator 
dynamics (Weibull, 1995). In this case, the total number of players who use a given strategy changes only due 
to the increase or decrease of the strategy fitness, as described by functions defined in (6-7). 

In the absence of replication, $\delta=0$, there is a conservation of the number 
of individuals playing each strategy in the whole population. This corresponds to our model 
assumption that individuals can not change their strategies but only habitats in which they live. 

For $f^1_A > f^2_A$ we obtain from (2-5) equations for $r_{i}(t)$ and $r(t)$:

\begin{equation}
r_{1}(t+\delta)=(1-\delta)r_{1}(t)+
\delta r_{1}(t)[\frac{r_{A}^{1}f_{A}^{1}+r_{B}^{1}f_{B}^{1}}{r_{1}}+
\gamma\frac{r^{2}_{A}(f_{A}^{1}-f_{A}^{2})+r^{1}_{B}(f_{B}^{1}-f_{B}^{2})}{r_{1}}],
\end{equation}

\begin{equation}
r_{2}(t+\delta)=(1-\kappa \delta)r_{2}(t)+
\delta r_{2}(t)[\kappa \frac{r_{A}^{2}f_{A}^{2}+r_{B}^{2}f_{B}^{2}}{r_{2}}+
\gamma\frac{r^{2}_{A}(f_{A}^{2}-f_{A}^{1})+r^{1}_{B}(f_{B}^{2}-f_{B}^{1})}{r_{2}}],
\end{equation}

\begin{equation} 
r(t+\delta)= (1-\delta)r_{1}(t)+ (1-\kappa \delta)r_{2}(t)  + 
\delta r(t)[\alpha ({\frac{r^1_A}{r_1}}f_{A}^{1}+ {\frac{r^1_B}{r_1}} f_{B}^{1})+
(1-\alpha)\kappa ( {\frac{r^2_A}{r_2}} f_{A}^{2}+ {\frac{r^2_B}{r_2}} f_{B}^{2})],
\end{equation}

where all functions in square brackets depend on $t$. 

Now, as in the derivation of standard replicator dynamics (cf. e.g. Weibull, 1995), 
we consider the frequencies of the individuals playing the relevant strategies in both 
habitats rather than their numbers. 
Thus, we focus on the temporal evolution of the frequencies, $x$ and $y$, and the relative 
size of the first subpopulation, $\alpha$. We divide (2) by (12), (4) by (13), and (12) by (14).
 Performing the limit $\delta \rightarrow 0$ we obtain the following differential equations:

\begin{equation}
\frac{dx}{dt}=x[(1-x)(f_{A}^{1}-f_{B}^{1})+\gamma[(\frac{y(1-\alpha)}{x\alpha}-\frac{y(1-\alpha)}{\alpha})
(f_{A}^{1}-f_{A}^{2})-(1-x)(f_{B}^{1}-f_{B}^{2})]],
\end{equation}

\begin{equation}
\frac{dy}{dt}=y[\kappa(1-y)(f_{A}^{2}-f_{B}^{2})+\gamma[(1-y)(f_{A}^{2}-f_{A}^{1})-
\frac{(1-x)\alpha}{1-\alpha}(f_{B}^{2}-f_{B}^{1})]],
\end{equation}

\begin{equation}
\frac{d\alpha}{dt}=\alpha(1-\alpha)[xf_{A}^{1}+(1-x)f_{B}^{1}-(yf_{A}^{2}+(1-y)f_{B}^{2})]
\end{equation}
$$+\alpha \gamma[\frac{y(1-\alpha)}{\alpha}(f_{A}^{1}-f_{A}^{2})+(1-x)(f_{B}^{1}-f_{B}^{2})]
+\alpha(1-\alpha)(\kappa-1)(1-yf^{2}_{A}-(1-y)f^{2}_{B}).$$
  
Similar equations are derived for the case $f^1_A < f^2_A$ (since our model is symmetric 
with respect to the permutation of the subpopulations, it is enough to renumerate the 
relevant indices and redefine the parameter $\kappa$). 

Assume first that $f_{A}^{1}(0)>f_{A}^{2}(0)$, which is equivalent to  
$x(0)>y(0)$. It follows from (2-5) that a fraction of $A$-players from the subpopulation $2$
will migrate to the subpopulation $1$ and a fraction of $B$-players will migrate in the opposite direction.
This will cause $x$ to increase and $y$ to decrease. However, if $x(0)<x^{*}$ and $y(0)<x^{*}$, 
then $f_{A}^{1}<f_{B}^{1}$ and $f_{A}^{2}<f_{B}^{2}$, therefore $B$-players will have more 
offspring than $A$-players. This has the opposite effect on the relative number of $A$-players 
in the first subpopulation than migration. If $x(0)<y(0)$, then migration takes place 
in the reverse directions.

The outcome of the competition between migration and replication depends, for a given payoff matrix, 
on the relation between $x(0)-y(0)$, $\gamma$ and $\kappa$. 
We are interested in formulating sufficient conditions for the parameters of the model, for which 
most individuals of the whole population will play in the long run the efficient strategy $A$. 
We prove the following theorem.

\vspace{2mm}

\noindent {\bf Theorem} 
\vspace{3mm} 

If 
$$\gamma [ x(0)-y(0) ] > max [\frac {d-b} {d-c}, \frac {\kappa (a-c)}{a-b}],$$
then $x(t) \rightarrow_{t \rightarrow \infty}1$ and $y(t) \rightarrow_{t \rightarrow \infty} 0$.
If $\kappa<(a-1)/(d-1)$, then $\alpha(t) \rightarrow_{t \rightarrow \infty} 1$. 
\vskip 0.2cm
If 
$$\gamma [ y(0)-x(0) ] > max [\frac {\kappa (d-b)} {d-c}, \frac {a-c} {a-b}],$$
then $x(t) \rightarrow_{t \rightarrow \infty}0$ and $y(t) \rightarrow_{t \rightarrow \infty} 1$.
If $\kappa > (d-1)/(a-1)$, then $\alpha(t) \rightarrow_{t \rightarrow \infty} 0.$ 
\vspace{2mm}

{\bf Proof}: 

Assume first that $x(0)>y(0)$. From (15-16) we get the following differential inequalities:

\begin{equation}
\frac{dx}{dt} > x(1-x)[f_{A}^{1}-f_{B}^{1})+\gamma(f_{B}^{2}-f_{B}^{1})],
\end{equation}

\begin{equation}
\frac{dy}{dt} < y(1-y)[\kappa(f_{A}^{2}-f_{B}^{2})+\gamma(f_{A}^{2}-f_{A}^{1})],
\end{equation}

Using explicit expressions for $f_{s}^{i}$ we get

\begin{equation}
\frac{dx}{dt} > x(1-x)[(a-c+d-b)x+b-d+\gamma(d-c)(x-y)], 
\end{equation}

\begin{equation}
\frac{dy}{dt} < y(1-y)[\kappa[(a-c+d-b)y+b-d]-\gamma(a-b)(x-y)],
\end{equation}

We note that if $\gamma(d-c)(x(0)-y(0)) > d - b$  
then $\gamma(d-c)(x(0)-y(0))+ b - d + (a-c+d-b)x(0) > 0,$ i.e. $dx/dt(0)>0$.  

Analogously, if $\gamma(a-b)(x(0)-y(0))> \kappa (a-c)$, then 
$\gamma(a-b)(x(0)-y(0))> \kappa [(a-c+d-b) + b-d] > \kappa [(a-c+d-b)y(o) + b-d],$ 
therefore $dy/dt(0)<0$. 
Thus, combining both conditions we conclude that $x(t)-y(t)$ is an increasing function so
$x(t)>y(t)$ for all $t \geq 0$, hence we may use (15-17) all the time. 
We get that
$x(t) \rightarrow_{t \rightarrow \infty}1$ and $y(t) \rightarrow_{t \rightarrow \infty} 0$, 
and the first part of the thesis follows. 
Now from (17) it follows that if $a-d+(\kappa-1)(1-d)>0$, i.e. $\kappa< (a-1)/(d-1)$, 
then $\alpha(t) \rightarrow_{t \rightarrow \infty} 1$. 

The second part of our Theorem, corresponding to initial conditions $y(0)>x(0),$ can be proved analogously, 
starting from eqs. (2-5) written for the case $f^1_A(0) < f^2_A(0)$ and their continuous 
counterparts. We omit details. $\Box$

The above conditions for $\kappa$ mean that the population consisting of just $A$-players 
replicates faster [exponentially in $(a-1)t$] than the one consisting of just $B$-players 
[exponentially in $(d-1)\kappa t$]. The same results would follow if the coefficients of the 
payoff matrix of the game played in one habitat would differ from those in the 
second habitat by an additive constant.
 
We have shown that for $\gamma$ and the difference of the initial 
conditions $x(0)-y(0)$ satisfying certain inequalities,  
the fraction of $A$-players in one of the subpopulations converges to $1$,  
the relative size of this subpopulation converges to $1$, 
and therefore the fraction of the whole population playing $A$, 
$z= r_{A}/r = x \alpha + y(1-\alpha)$ converges to $1$. Which subpopulation prevails in this sense 
depends on the initial conditions, whether $x(0)>y(0)$ or $y(0)>x(0)$.

Let us note that the main result of our paper, as stated in the Theorem, 
is also valid in the special case of equal replicator rates in each habitat,
i.e. when $\kappa =1$. 

\section{Conclusions}

We discussed a dynamics of a single population of individuals playing 
a symmetric game. At any time, each individual resides in one 
of two disjoint different habitats. Differences between habitats 
(e.g. different ecology, fauna, flora, and feeding capacities) 
result in different speeds of replication. Players have knowledge 
of the expected payoffs from the strategy they are identified with in both habitats, 
and they choose the one with a higher expected payoff. The process of changing a habitat 
is referred to as migration. We assumed that the coefficient of migration is the same 
in both subpopulations. More general case would give another (apart of the difference 
in replicator speeds) characterization of both habitats. 

We showed that introduction of this  mechanism of ``attraction'' by the habitat with a higher expected payoff 
in the standard replicator dynamics can help the whole population to reach the state 
in which in the long run most individuals play the efficient strategy. 

More precisely, we proved that for a given rate of migration, if the fractions of individuals 
playing the efficient strategy in both habitats are not too close to each other, then the habitat 
with a higher fraction of such players overcomes the other one in the long run. 
The fraction of individuals playing the efficient strategy tends to unity in this habitat 
and consequently in the whole population. Alternatively, we may say that the bigger 
the rate of migration is, larger is the basin of attraction of the efficient equilibrium. 
In particular, we showed that for a large range of parameters of our dynamics,
even if the initial conditions in both habitats are in the basin 
of attraction of the risk-dominant equilibrium (with respect 
to the standard replication dynamics without migration), 
in the long run most individuals play the efficient strategy.

The sizes of basins of attraction play an important role when deterministic dynamics is subject 
to stochastic perturbations due to mutations or players mistakes. The long-run behavior 
of noisy evolutionary processes can be described by stochastic stability of deterministic equilibria 
(Foster and Young, 1990). It was shown in many models that equilibria with larger basins of attraction 
(risk-dominant ones in the standard replicator dynamics) are stochastically stable 
(Harris and Fudenberg, 1992; Kandori {\em et al.}, 1993); see also Robson and Vega-Redondo (1996), 
Vega-Redondo (1996) and Mi\c{e}kisz (2005) for models with randomly matched players. 
This is a consequence of the fact that the population needs more mutations to evolve 
from an equilibrium with a larger basin of attraction to the other equilibrium 
than in the case of the opposite transition. Therefore, for a low level of mutations, 
the population settles in the long run with a high probability in a state 
with a larger basin of attraction.    

We plan to investigate in the future an appropriate stochastic version of our model.
One has to carefully examine basins of attraction of both equilibria. The result of our paper 
suggests that the efficient equilibrium would be stochastically stable.
\vspace{3mm}

\noindent {\bf Acknowledgments}
\vspace{3mm}

\noindent We thank the Polish Committee for Scientific Research 
for financial support under the grant KBN 5 P03A 025 20
and an anonymous referee for helpful suggestions.

\vspace{3mm}

\noindent {\bf References}
\vspace{4mm}

\noindent Foster, D., Young, P. H., 1990.
Stochastic evolutionary game dynamics.
{\em Theoretical Population Biology} 38, 219-232.
\vspace{2mm}

\noindent Harris, C., Fudenberg, D., 1992. Evolutionary dynamics with aggregated shocks. 
{\em J. Econ. Theory} 57, 420-441.
\vspace{2mm}

\noindent Hars\'{a}nyi, J. and Selten, R., 1988.
{\em A General Theory of Equilibrium Selection in Games.} 
MIT Press, Cambridge MA. 
\vspace{2mm}

\noindent Hofbauer, J., Schuster, P., Sigmund, K., 1979.
A note on evolutionarily stable strategies and game dynamics.
{\em J. Theor. Biol.} 81, 609-612.
\vspace{2mm}

\noindent Hofbauer, J., Sigmund, K., 1998. 
{\em Evolutionary Games and Population Dynamics.}
Cambridge University Press.
\vspace{2mm}
 
\noindent Hofbauer, J., Sigmund, K., 2003. Evolutionary game dynamics, 
{\em Bulletin AMS} 40, 479-519. 

\noindent Kandori, M., Mailath G. J., Rob, R., 1993.
Learning, mutation, and long-run equilibria in games. 
{\em Econometrica} 61, 29-56. 
\vspace{2mm}

\noindent Maynard Smith, J., Price G. R., 1973. The logic of animal conflicts. 
{\em Nature} 246, 15-18.
\vspace{2mm}

\noindent Maynard Smith, J., 1982. {\em Evolution and the Theory of Games}.
Cambridge University Press.
\vspace{2mm}

\noindent Mi\c{e}kisz, J., 2005. Equilibrium selection in evolutionary games 
with random matching of players, {\em J. Theor. Biol.} 232, 47-53. 
\vspace{2mm}

\noindent Robson, A., Vega-Redondo, F., 1996. Efficient equilibrium selection
in evolutionary games with random matching. {\em J. Econ. Theory} 70, 65-92.
\vspace{2mm}

\noindent Taylor, P. D., Jonker, L. B., 1978. Evolutionarily stable
strategy and game dynamics. {\em Math. Biosci.} 40, 145-156.
\vspace{2mm}

\noindent Vega-Redondo, F., 1996. {\em Evolution, Games, and Economic Behaviour.}
Oxford University Press.
\vspace{2mm}

\noindent Weibull, J., 1995. {\em Evolutionary Game Theory.} MIT Press, Cambridge MA.
\vspace{2mm}

\noindent Zeeman, E., 1981. Dynamics of the evolution of animal conflicts.
{\em J. Theor. Biol.} 89, 249-270.

\end{document}